\documentclass[jmp,graphicx]{revtex4-1}

\usepackage{ amssymb }
\usepackage{amsmath}
\usepackage{breqn}
\draft 

\begin{document}


\title{Solving Linear Differential Equations by recursion and integrating factors} 



\author{Everardo Rivera-Oliva}
\email{everardo.rivera@cinvestav.mx}

\affiliation{{ 
Departamento de Física\\
Centro de Investigación y de Estudios Avanzados del Instituto Politécnico Nacional\\
P.O. box 14-740 C.P. 07000 Ciudad de México;México}
}


\date{\today}

\begin{abstract}
In this study, a recursive solution technique in conjunction with generalized integrating factors is presented and applied to address first and second order linear differential equations. This approach demonstrates practical utility in classical differential equations encountered in physics, inclusive of equations with variable coefficients, particularly when a pattern within the recursion is identifiable, thus enabling the derivation of an explicit expression for $y(x)$.
\end{abstract}

\maketitle 


\section{Introduction}
Numerous problems encountered in physics are expressed through differential equations, a prevalent framework utilized for studying various phenomena. Typically, these differential equations in physics are either first or second order and may be categorized as either ordinary or partial differential equations. For the case of partial differential equations, predominant methodologies such as separation of variables(see \cite{arfken2013mathematical} for a review) or the application of Green's functions (see \cite{butkov1968mathematical} for a review) typically aim to decompose the partial differential equation into a series of ordinary differential equations or to transform the partial differential equation into an algebraic problem. Regardless, when addressing partial differential equations, it is often necessary to resolve a system of ordinary differential equations.\\
Many of the prevalent ordinary differential equations encountered in physics are linear in nature and can be broadly categorized into two groups: those with constant coefficients and those with variable coefficients. Regardless of the category, traditional approaches to finding solutions rely heavily on conjecture. In the scenario of constant coefficients, a proposed solution takes the form of $\exp(\lambda x)$, where $\lambda$ is a constant determined by imposing the solution's conformity with the differential equation. For equations with variable coefficients, the most commonly employed techniques involve the transformation of variables (see \cite{zill2016differential}), aiming to alter the independent variable such that the equation can be transformed into one with constant coefficients; this is notably the approach for addressing the Cauchy-Euler equation (see \cite{zill2016differential}). Another frequently utilized method is the Frobenius method (see \cite{butkov1968mathematical} for a review), wherein a solution is posited as a series expansion in terms of the independent variable, and through substitution into the differential equation, a recurrence relation is ascertained for calculating the series coefficients. By resolving this relation and determining each coefficient, a solution to the differential equation can be constructed. In both scenarios, when an inhomogeneous term is present, a particular solution is again deduced via guessing through the method of undetermined coefficients or by employing the variation of parameters method. Therefore, resolving second-order linear differential equations predominantly involves conjectural approaches. Conversely, for first-order linear differential equations, the integrating factor method allows for a systematic, logical procedure to be employed in deriving the solution, obviating the need for speculative solution forms. However, for second-order cases, no analogous systematic method exists, necessitating the proposal of specific solution forms.\\
This study investigates the development and implementation of a systematic logical approach for solving ordinary differential equations utilizing recursion. This recursive method can be integrated with existing solution techniques, facilitating the resolution of more complex differential equations. Although the methodology is applicable to nonlinear differential equations, this paper focuses on first and second order equations that frequently arise in the field of physics. It will be demonstrated that recursion directly enables the derivation of general solutions for first-order linear differential equations in both homogeneous and inhomogeneous instances. For second-order linear differential equations, a systematic solution technique is derived, inspired by the integrating factor method for first-order equations. By identifying two integrating factors $\alpha(x),\beta(x)$ through recursion, the differential equation is transformed into an exact equation. This transformation reduces the problem to a first-order linear equation, which, upon integration, facilitates the reconstruction of the general solution for the second-order linear differential equation.
\\The organization of this work is as follows: Section \ref{2} introduces the recursive method for solving ordinary differential equations. Section \ref{3} applies this method to the general first-order differential equation, providing a comparison with the existing integrating factor method. In Section \ref{4}, a generalized method of integrating factors is developed, utilizing recursion to address second-order linear differential equations. Section \ref{5} demonstrates the application of the method to classical second-order linear differential equations encountered in physics. Finally, Section \ref{6} discusses conclusions, perspectives, and future directions.

\section{Recursive method to solve ordinary differential equations}\label{2}
As delineated in any conventional text on differential equations (for an introductory treatment, refer to \cite{zill2016differential}; for comprehensive reviews, see \cite{arfken2013mathematical}, \cite{butkov1968mathematical}, and \cite{hassani2013mathematical}), an ordinary differential equation concerning the unknown function $y(x)$ constitutes an equation encompassing $y(x)$ and the derivatives of $y(x)$ with respect to the independent variable $x$. Generally, an ordinary differential equation for $y(x)$ can be expressed as:
\begin{equation}\label{r1}
    F\left(y(x),\frac{dy}{dx},\frac{d^2y}{dx^2},\hdots,\frac{d^ny}{dx^n},x,f(x)\right)=0
.\end{equation}
where $f(x)$ is identified as the inhomogeneous function of the equation, which is independent of $y(x)$. The order of the differential equation delineated by Eq.\eqref{r1} is determined by the order of the highest derivative of $y(x)$ that is present within it. The classification of the existing types of ordinary differential equations depicted by Eq.\eqref{r1} is constructed upon the properties fulfilled by $F(y,y',...,y^{(n)},f)$.\\
A general solution method for differential equations of the form presented in Eq.\eqref{r1} does not exist; solutions are available only for particular instances of the functional form $F(y,y',...,y^{(n)},f)$ (see \cite{zill2016differential,arfken2013mathematical,butkov1968mathematical}). In this context, a recursive method is introduced to attempt solving differential equations of the form indicated by Eq.\eqref{r1}. This recursive method involves solving for $y(x)$ expressed in terms of $y(x)$, meaning that the solution $y(x)$ is articulated as a function of itself and its derivatives.The approach consists in reformulating $F(y,y',...,y^{(n)},f)$ into a known, solvable differential equation $H(y,y',...,y^{(m)},g)$ that is sufficiently simple to be solved for $y(x)$:
\begin{equation}
    F(y,y',...,y^{(n)},f)=H(y,y',...,y^{(m)},g)
.\end{equation}
where $H(y,y',...,y^{(m)},g)$ represents a differential equation for $y(x)$ that is known to be solvable. This is accomplished at the cost that the inhomogeneous function within $H(y,y',...,y^{(m)},g)$ is not independent from $y(x)$ or its derivatives, but is instead a function dependent upon them:
\begin{equation}
    g=g(y,y',..,y^{(k)},x)
.\end{equation}
Consequently, the solution for $y(x)$ derived via this method will be articulated as a function of $y(x)$ itself along with its derivatives, manifesting as a recursion defined by:
\begin{equation}\label{r2}
    y(x) = G\left(y(x),\frac{dy}{dx},\frac{d^2y}{dx^2},\hdots,\frac{d^ny}{dx^n},x,f(x)\right)
.\end{equation}
such that:
\begin{equation}\label{r3}
    F\left(G,\frac{dG}{dx},\frac{d^2G}{dx^2},\hdots,\frac{d^nG}{dx^n},x,f(x)\right)=0
.\end{equation}
Equation \eqref{r2} is a recursive equation for $y(x)$ and constitutes a solution to the ordinary differential equation delineated in Eq.\eqref{r1}. By subsequently executing recursive substitutions of the recursive relationship for $y(x)$ into $y(x)$, one can find the explicit solution for Eq.\eqref{r1}:
\begin{equation}
    y(x)=G(G(G...(G(x))...))
.\end{equation}
In general, the recursive relationship may necessitate an infinite number of iterations, with no assurance of convergence to a general solution or convergence of the function at all. The method exhibits practical utility when a pattern within the recursion can be discerned, thereby facilitating the derivation of an explicit expression for $y(x)$. A notable strength of this approach lies in its ability to be employed in conjunction with other established methodologies for solving differential equations. In the present work, it will be utilized in tandem with additional techniques to address second order ordinary differential equations.
\section{First Order Linear Equations}\label{3}
The examination begins with First Order Ordinary Differential Equations. The canonical form of a first-order inhomogeneous linear differential equation is represented as follows:

\begin{equation}\label{eq 1.1}
    \frac{dy}{dx}+p(x)y(x)=f(x)
.\end{equation}
The Integrating Factor Method, which is a widely recognized approach to solving Eq.\eqref{eq 1.1}, is comprehensively documented in the literature. For an initial introduction, refer to \cite{zill2016differential,arfken2013mathematical,butkov1968mathematical,hassani2013mathematical}. This method involves multiplying Eq.\eqref{eq 1.1} by an integrating factor function $\mu(x)$, enabling the transformation of Eq.\eqref{eq 1.1} into an exact differential equation. By applying this factor to Eq.\eqref{eq 1.1}, it is derived the following equation:

\begin{equation}\label{eq 1.2}
    \mu(x)\frac{dy}{dx}+\mu(x)p(x)y(x)=\mu(x)f(x)
.\end{equation}
The objective is to ensure that the left-hand side of Eq.\eqref{eq 1.2} is represented as an exact differential, denoted by:

\begin{equation}\label{eq 1.3}
    \frac{d}{dx}(\mu(x)y(x)) = \mu(x)\frac{dy}{dx}+\frac{d\mu}{dx}y(x)
.\end{equation}
To ensure the equivalence of Eq.\eqref{eq 1.3} and Eq.\eqref{eq 1.2}, it is necessary for the following condition to be fulfilled:

\begin{equation}\label{eq 1.4}
    \frac{d\mu}{dx}=\mu(x)p(x)
.\end{equation}
By solving Equation \eqref{eq 1.4}, it is derived the expression for the integrating factor $\mu(x)$:
\begin{equation}\label{eq 1.5}
    \mu(x) = \exp\left({\int_0^x p(t)dt}\right)
.\end{equation}
Incorporating the expression for $\mu(x)$ as delineated by Eq.\eqref{eq 1.5} into Eq.\eqref{eq 1.2}, an exact differential equation for $y(x)$ can thus be reformulated as follows:
\begin{equation}\label{eq 1.6}
    \frac{d}{dx} \left(y(x)  \exp\left({\int_0^x p(t)dt}\right)\right)= \exp\left({\int_0^x p(t)dt}\right)f(x)
.\end{equation}
Through direct integration of Eq.\eqref{eq 1.6}, it is obtained the general solution for $y(x)$ presented as:
\begin{equation}\label{eq 1.7}
    y(x) = C_1 \exp \left({-\int_0^x p(t)dt}\right)+ \exp\left({-\int_0^x p(t)dt}\right)\int_0^x  \exp \left({\int_0^t p(t')dt'}\right)f(t)dt
.\end{equation}
Equation \eqref{eq 1.7} delineates the general solution to the ordinary differential equation articulated by Eq.\eqref{eq 1.1}, as documented in any standard differential equations textbook (see \cite{zill2016differential,arfken2013mathematical,butkov1968mathematical}).

\subsection{Recursive solution method}
An attempt to solve Eq.\eqref{eq 1.1} is made utilizing the recursive methodology delineated in Section \ref{2}. As outlined, this technique entails resolving $y(x)$ in terms of $y(x)$ and its derivatives. A crucial component, as previously indicated, is to recast the ordinary differential equation (ODE) into a form that corresponds to a known solvable ODE. Specifically, for the ODE presented in Eq.\eqref{eq 1.1}, the solution for the first derivative of $y(x)$ is directly obtainable:
\begin{equation}\label{eq 1.2.1}
    \frac{dy}{dx}=g(x)=f(x)-p(x)y(x)
.\end{equation}
In this scenario, the linear first-order differential equation is reformulated into an equation of the form $y'=g$, wherein $g=f(x)-p(x)y(x)$ serves as an inhomogeneous function of $y(x)$ itself. Subsequently, Eq.\eqref{eq 1.2.1} is integrated to obtain:
\begin{equation}\label{eq 1.2.2}
    y(x)=C_1 +\int_0^x \left(f(t)-p(t)y(t) \right)dt
.\end{equation}
What has been done is to treat the ODE as if the only $y(x)$ dependence were through a first derivative and treating $p(x)y(x)$ as an extra inhomogeneous term at the same level of $f(x)$. By doing so, it was obtained Eq.\eqref{eq 1.2.2} which is a recursive solution for $y(x)$. This is a Volterra equation of the second kind. The subsequent task involves resolving the recursion by substituting Eq.\eqref{eq 1.2.2} into Eq.\eqref{eq 1.2.2}.Upon executing the first substitution it is obtained:

\begin{equation}\label{eq 1.2.3}
\begin{split}
    y(x)=&\int_0^x f(t_1)dt_1-\int_0^x \int_0^{t_1}p(t_1)f(t_2)dt_2dt_1+\int_0^x\int_0^{t_1}p(t_1)p(t_2)y(t_2)dt_2dt_1\\&+C_1-C_1 \int_0^x p(t_1)dt_1.
\end{split}
\end{equation}
Through the application of a subsequent recursive step, it is derived:
\begin{equation}\label{eq 1.2.4}
\begin{split}
    y(x)=&\int_0^x f(t_1)dt_1-\int_0^x \int_0^{t_1}p(t_1)f(t_2)dt_2dt_1+\int_0^x\int_0^{t_1}\int_0^{t_2}p(t_1)p(t_2)f(t_3)dt_3dt_2dt_1\\&+C_1-C_1 \int_0^x p(t_1)dt_1+C_1 \int_0^x \int_0^{t_1} p(t_1)p(t_2)dt_2dt_1+\hdots 
\end{split}
\end{equation}
An examination of Equations \eqref{eq 1.2.3} and \eqref{eq 1.2.4} reveals a discernible pattern inherent in the recursive solution:

\begin{equation} \label{eq 1.2.5}
\begin{split}
    y(x)=&\int_0^xf(t_1)dt_1+\int_0^x \sum_{i=1}^{\infty} (-1)^i \prod_{j=1}^{i}\int_0^{t_j}p(t_j)f(t_{j+1})dt_{j+1}dt_{j}\\&+C_1\sum_{i=1}^{\infty}(-1)^{i}\int_0^x \prod_{j=1}^{i} \int_0^{t_j}p(t_j)dt_j+C_1.
    \end{split}
\end{equation}
The integrals present within the $C_1$ terms can be articulated as follows:

\begin{equation}\label{eq 1.2.6}
    \int_0^x \prod_{j=1}^{i} \int_0^{t_j}p(t_j)dt_j = \frac{1}{i!}\left[ \int_0^x p(t)dt\right]^{i}
.\end{equation}
Then Eq.\eqref{eq 1.2.5} reduces to:
\begin{equation}\label{eq 1.2.7}
    y(x)=C_1\sum_{i=0}^{\infty} \frac{(-1)^{i}}{i!}\left[ \int_0^x p(t)dt\right]^{i}+\int_0^xf(t_1)dt_1+\int_0^x \sum_{i=1}^{\infty} (-1)^i \prod_{j=1}^{i}\int_0^{t_j}p(t_j)f(t_{j+1})dt_{j+1}dt_{j}.
\end{equation}
This expression corresponds exactly to the Taylor series expansion for an exponential function:
\begin{equation}\label{eq 1.2.8}
    \sum_{i=0}^{\infty} \frac{(-1)^{i}}{i!}\left[ \int_0^x p(t)dt\right]^{i}= \exp\left({-\int_0^x p(t)dt}\right)
.\end{equation}

Consequently, $y(x)$ can be reformulated as follows:
\begin{equation}\label{eq 1.2.9}
    y(x)=C_1\exp \left({-\int_0^x p(t)dt}\right)+\int_0^xf(t_1)dt_1+\int_0^x \sum_{i=1}^{\infty} (-1)^i \prod_{j=1}^{i}\int_0^{t_j}p(t_j)f(t_{j+1})dt_{j+1}dt_{j}
.\end{equation}

The recursive integrals presented in Eq.\eqref{eq 1.2.9} can be systematically disentangled. As an illustration, consider the term:

\begin{equation}\label{eq 1.2.10}
    \int_0^x \int_0^{t_1}\int_0^{t_2}p(t_1)p(t_2)f(t_3)dt_3dt_2dt_1= \int_0^x \frac{1}{2!}\left(\int_t^x (-1)p(t')dt'\right)^2f(t) dt
.\end{equation}
This procedure can be applied to each term presented in Eq.\eqref{eq 1.2.9}, thereby simplifying the expression for $y(x)$ to:

\begin{equation}\label{eq 1.2.11}
    y(x)=C_1\exp \left({-\int_0^x f(t)dt}\right)+\int_0^x \exp \left({-\int_t^x p(t')dt'}\right)f(t)dt
.\end{equation}
But:
\begin{equation}\label{eq 1.2.12}
    -\int_t^x p(t)dt=\int_0^t p(t')dt'-\int_0^x p(t)dt
.\end{equation}
Thus, the formulation of the recursive solution can ultimately be expressed as:
\begin{equation}\label{eq 1.2.13}
    y(x)=C_1\exp \left({-\int_0^x p(t)dt}\right)+\exp \left({-\int_0^x p(t)dt}\right)\int_0^x f(t)\exp\left({\int_0^t p(t')dt'}\right)dt
.\end{equation}
Evidently, the recursive solution converges to the same expression derived using the integrating factor method, as presented in Eq. \eqref{eq 1.7}. This demonstrates that the recursive solution method is effective in determining the general solution for first-order linear ordinary differential equations.

\section{Second Order Linear Equations:Generalized recursive integrating factors}\label{4}
The exploration of second-order linear differential equations will now proceed. The standard form of a second-order ODE is given by:
\begin{equation} \label{eq 2.1.1}
    \frac{d^2y}{dx^2}+p(x)\frac{dy}{dx}+q(x)y(x)=f(x)
.\end{equation}
Methods to solve Eq.\eqref{eq 2.1.1} are applicable based on the conditions that $p(x),q(x),f(x)$ fulfill. In the field of physics, one of the most commonly employed techniques is the Frobenius method delineated in \cite{+1873+214+235} concerning series solutions. Nonetheless, for specific differential equations, such as those with constant coefficients and Cauchy-Euler equations, particular methods exist as those present in \cite{zill2016differential,arfken2013mathematical,butkov1968mathematical}.
The primary advantage of the recursive method lies in its capacity to transform ordinary differential equations into known solvable forms, enabling integration into well-understood expressions. Additionally, this method can be synergistically combined with other solution techniques when deemed beneficial. Consequently, the essence of the integrating factor method pertinent to first-order equations is adopted to reformulate Eq.\eqref{eq 2.1.1} as an exact differential equation by multiplying Eq.\eqref{eq 2.1.1} by two functions $\alpha(x),\beta(x)$:
\begin{equation} \label{eq 2.1.2}
    \alpha(x)\beta(x)\left(\frac{d^2y}{dx^2}+p(x)\frac{dy}{dx}+q(x)y(x)\right)=f(x) \alpha(x)\beta(x)
.\end{equation}
Observe that:
\begin{equation} \label{eq 2.1.3}
    \frac{d}{dx}\left(\alpha(x)\beta(x)\frac{dy}{dx}\right)=\alpha(x)\beta(x)\frac{d^2y}{dx^2}+\frac{d\alpha}{dx}\beta(x)\frac{dy}{dx}+\alpha(x)\frac{d\beta}{dx}\frac{dy}{dx}
.\end{equation}

\begin{equation}\label{eq 2.1.4}
    \frac{d}{dx}\left(\frac{d\alpha}{dx}\beta(x)y(x) \right) = \frac{d^2\alpha}{dx^2}\beta(x)y(x)+\frac{d\alpha}{dx}\frac{d\beta}{dx}y(x)+\frac{d\alpha}{dx}\beta(x)\frac{dy}{dx}
.\end{equation}
\begin{equation}\label{eq 2.1.5}
    \frac{d}{dx}\left(\alpha(x)\frac{d\beta}{dx}y(x)\right)=\frac{d\alpha}{dx}\frac{d\beta}{dx}y(x)+\alpha \frac{d^2 \beta}{dx^2}y(x)+\alpha(x)\frac{d\beta}{dx}\frac{dy}{dx}
.\end{equation}
By adding Eq. \eqref{eq 2.1.3} to \eqref{eq 2.1.5} and subsequently subtracting Eq. \eqref{eq 2.1.4}, it is derived the following:
\begin{equation}\label{eq 2.1.6}
\begin{split}
    \frac{d}{dx}\left(\alpha(x)\beta(x)\frac{dy}{dx}\right)- \frac{d}{dx}\left(\frac{d\alpha}{dx}\beta(x)y \right)+\frac{d}{dx}\left(\alpha(x)\frac{d\beta}{dx}y\right) = \alpha(x) \beta(x)\frac{d^2y}{dx^2}+2\alpha(x)\frac{d\beta}{dx}\frac{dy}{dx}+\alpha(x)\frac{d^2\beta}{dx^2}y-\beta(x)\frac{d^2\alpha}{dx^2}y.
\end{split}
\end{equation}
In order for Eq.\eqref{eq 2.1.2} to be transformed into an exact differential, it is necessary that the following conditions are satisfied:
\begin{equation}\label{eq 2.1.7}
    \frac{d\beta}{dx}=\beta(x)\frac{p(x)}{2}
.\end{equation}
\begin{equation}\label{eq 2.1.8}
    \alpha(x)\frac{d^2\beta}{dx^2}-\beta(x)\frac{d^2\alpha}{dx^2}=\alpha(x)\beta(x)q(x)
.\end{equation}
Should $\alpha(x),\beta(x)$ be identified in compliance with these stipulated conditions, a first integral can subsequently be constructed:
\begin{equation} \label{eq 2.1.9}
    \alpha(x)\beta(x)\frac{dy}{dx}+\left(-\frac{d\alpha}{dx}\beta(x)+\alpha(x)\frac{d\beta}{dx}\right)y(x)= \int_0^x \alpha(t)\beta(t)f(t)dt + C_1
.\end{equation}
Subsequently, a first-order linear differential equation is derived. When this first-order ODE is restructured into its canonical form, it is obtained:
\begin{equation}\label{eq 2.1.10}
    \frac{dy}{dx}+\frac{d}{dx}\ln\left( \frac{\beta}{\alpha}\right) y(x)=\frac{C_1}{\alpha(x)\beta(x)}+\frac{1}{\alpha(x) \beta(x)}\int_0^x\alpha(t)\beta(t)f(t)dt
.\end{equation}
The first-order ordinary differential equation presented in Eq.\eqref{eq 2.1.10} can be addressed using the integrating factor method or alternatively, the previously examined recursive method:
\begin{equation}\label{eq 2.1.11}
    \mu(x) = \frac{\beta(x)}{\alpha(x)}
.\end{equation}
then:
\begin{equation}\label{eq 2.1.12}
    \frac{d}{dx} \left(y(x)\frac{\beta}{\alpha(x)} \right)= \frac{C_1}{\alpha^2(x)}+\frac{1}{\alpha^2(x)}\int_0^x \alpha(t)\beta(t)f(t)dt
.\end{equation}
Consequently, integration of this equation yields $y(x)$:\begin{equation}\label{eq 2.1.13}
    y(x)=C_2\frac{\alpha(x)}{\beta(x)}+C_1\frac{\alpha(x)}{\beta(x)}\int_0^x \frac{dt}{\alpha^2(t)}+\frac{\alpha(x)}{\beta(x)}\int_0^x \frac{1}{\alpha^2(t)}\int_0^t \alpha(t')\beta(t')f(t')dt' dt
.\end{equation}
This constitutes the general solution to the second-order linear differential equation as indicated in Eq.\eqref{eq 2.1.1}. Hence, provided that $\alpha(x),\beta(x)$ is identifiable, it is possible to determine the general solution for $y(x)$ by implementing this methodology, to be named generalized recursive integrating factors.
Solving Eq.\eqref{eq 2.1.7} $\beta(x)$ is obtained:
\begin{equation}\label{eq 2.1.14}
    \beta(x)=\exp\left({\int_0^x \frac{p(x)}{2}}\right)
.\end{equation}
By substituting Eq.\eqref{eq 2.1.7} into Eq.\eqref{eq 2.1.8}, one obtains an equation for $\alpha(x)$:
\begin{equation} \label{eq 2.1.15}
    \alpha(x)\left(\frac{1}{2}\frac{dp}{dx}+\frac{p^2}{4} \right)-\frac{d^2\alpha}{dx^2}=\alpha(x)q(x)
.\end{equation}
then in canonical form:
\begin{equation}\label{eq 2.1.16}
    \frac{d^2\alpha}{dx^2}+\alpha(x) \left[q(x)-\frac{1}{2}\frac{dp}{dx}-\frac{p^2}{4} \right]=0
.\end{equation}
Define $h(x)$ by:
\begin{equation}\label{eq 2.1.17}
    h(x)=-q(x)+\frac{1}{2}\frac{dp}{dx}+\frac{p^2}{4}
.\end{equation}
Consequently Eq.\eqref{eq 2.1.16} can be rewritten as:
\begin{equation}\label{eq 2.1.18}
\frac{d^2 \alpha}{dx^2}-h(x)\alpha(x)=0
.\end{equation}
To determine $\alpha(x)$, it is necessary to solve a second-order linear ordinary differential equation, for which the general solution is not typically known. Accordingly, it shall be utilized the recursion method, directly addressing the equation's second derivative while treating all other terms as inhomogeneous components. Then a first integral for $\alpha(x)$ is obtained:
\begin{equation}\label{eq 2.1.19}
    \alpha(x) =C_2+C_1x+\int_0^x \int_0^{t_1} h(t_2)\alpha(t_2)dt_2dt_1
.\end{equation}
The aim is not to determine the general solution $\alpha(x)$, but rather to identify a particular solution. For the sake of simplicity, assume $C_1=0,C_2=1$. Consequently:
\begin{equation}\label{eq 2.1.20}
    \alpha(x)=1+\int_0^x \int_0^{t_1}h(t_2)\alpha(t_2)dt_2dt_1
.\end{equation}
By working out the first recursion it is obtained:
\begin{equation}\label{eq 2.1.21}
    \alpha(x)=1+\int_0^x\int_0^{t_1}h(t_2)dt_2dt_1+\int_0^x \int_0^{t_1}\int_0^{t_2}\int_0^{t_3}h(t_2)\alpha(t_4)dt_4dt_3dt_2dt_1
.\end{equation}
At this juncture, an evident pattern emerges within the recursive process:\begin{equation}\label{eq 2.1.22}
    \alpha(x)=1+\int_0^x\int_0^{t_1}h(t_2)dt_2dt_1+\int_0^x \int_0^{t_1}\int_0^{t_2}\int_0^{t_3}h(t_2)h(t_4)dt_4dt_3dt_2dt_1+\hdots
\end{equation}
In a compact notation $\alpha(x)$ can be written as:
\begin{equation}\label{eq 2.1.23}
\alpha(x)=1+\sum_{j\in 2 \mathbb{Z}^{+}}^{\infty}\prod_{i \in 2\mathbb{Z}^{+}}^{j} \int_0^{t_i}\int_0^{t_{i+1}} h(t_{i+2})dt_{i+2}dt_{i+1}
.\end{equation}
where $t_0=x$. It can be demonstrated that  Eq.\eqref{eq 2.1.22} satisfies Eq.\eqref{eq 2.1.18} by computing the second derivative of $\alpha(x)$:
\begin{equation}\label{eq 2.1.24}
    \frac{d\alpha}{dx} = \int_0^x h(t_2)dt_2+\int_0^{x}\int_0^{t_2}\int_0^{t_3}h(t_2)h(t_4)dt_4dt_3dt_2+\hdots
.\end{equation}
\begin{equation}\label{eq 2.1.25}
    \frac{d^2\alpha}{dx^2}=h(x)+h(x)\int_0^{x}\int_0^{t_3}h(t_4)dt_4dt_3+\hdots
.\end{equation}
\begin{equation}\label{eq 2.1.26}
    \frac{d^2\alpha}{dx^2}=h(x) \left[1+\int_0^x\int_0^{t_1}h(t_2)dt_2dt_1+\int_0^x \int_0^{t_1}\int_0^{t_2}\int_0^{t_3}h(t_2)h(t_4)dt_4dt_3dt_2dt_1+\hdots \right]
.\end{equation}
\begin{equation}\label{eq 2.1.27}
    \frac{d^2\alpha}{dx^2} = h(x)\alpha(x)
.\end{equation}
Consequently, $\alpha(x)$ as provided in Eq.\eqref{eq 2.1.21} serves as a solution to Eq.\eqref{eq 2.1.27} and, as a result, constitutes an integrating factor for Eq.\eqref{eq 2.1.1}. By reordering the integration sequence for $\alpha(x)$, simplification of the expression can be achieved:
\begin{equation}\label{eq 2.1.28}
    \alpha(x)=1+\int_0^x(x-t)h(t)dt+\int_0^x\int_0^t(x-t)(t-t_1)h(t)h(t_1)dt_1dt+\hdots
\end{equation}
By defining $x=t_0$ and definining a function $H(x)$ by:
\begin{equation}\label{eq 2.1.29}
    H(t_n)=(t_{n-1}-t_n)h(t_n)
.\end{equation}
then $\alpha(x)$ can be rewritten as:
\begin{equation}\label{eq 2.1.30}
    \alpha(x) =1+\int_0^{t_0} H(t_1)dt_1+\int_0^{t_0} \int_0^{t_1} H(t_2)H(t_1)dt_2dt_1+\hdots
\end{equation}
But Eq.\eqref{eq 2.1.30} is the expansion of the product ordered exponential (see \cite{grossman1972non} for a review). Consequently:
\begin{equation}
    \alpha(x) = \tau\left[\exp\left({\int_0^{x}H(t)dt}\right)\right]
.\end{equation}
Therefore $y(x)$ can be explicitly written as:

\begin{equation}\label{extra1}
\begin{split}
    y(x)=&C_2e^{-\int_{0}^{x} \frac{p(t)}{2}dt} \tau\left[e^{\int_0^{x}(x-t)\left(-q(t)+\frac{1}{2}\frac{dp}{dt}+\frac{p^2}{4} \right)dt}\right]\\ &+C_1e^{-\int_0^x \frac{p(t)}{2}dt} \tau\left[e^{\int_0^{x} (x-t)\left(-q(t)+\frac{1}{2}\frac{dp}{dt}+\frac{p^2}{4} \right)dt}\right] \int_0^x \frac{dt}{\tau\left[e^{-\int_0^{t}(t-t')\left(-q(t')+\frac{1}{2}\frac{dp}{dt'}+\frac{p^2}{4} \right)dt'}\right]^{2}}\\
    &+e^{-\int_0^x \frac{p(t)}{2}dt} \tau\left[e^{\int_0^{x} (x-t)\left(-q(t)+\frac{1}{2}\frac{dp}{dt}+\frac{p^2}{4} \right)dt}\right] \int_0^x\frac{\int_0^t f(t')e^{\int_0^{t'} \frac{p(z)}{2}dz} \tau\left[e^{\int_0^{t'}(t'-z)\left(-q(z)+\frac{1}{2}\frac{dp}{dz}+\frac{p^2}{4} \right)dz}\right]}{\tau\left[e^{-\int_0^{t}(t-t')\left(-q(t')+\frac{1}{2}\frac{dp}{dt'}+\frac{p^2}{4} \right)dt'}\right]^{2}}dt'dt.
\end{split}
\end{equation}
Equation \eqref{extra1} delineates the general solution to the second-order linear differential equation as articulated by Eq. \eqref{eq 2.1.1}. Although the notation may appear cumbersome in contrast to the succinct representation of Eq. \eqref{eq 2.1.12} with regard to integrating factors, it confers the benefit of being articulated explicitly in terms of the functions $p(x),q(x)$ present within the differential equation. Consequently, through the application of the generalized integrating factor method in tandem with the recursive method, second-order linear differential equations can be addressed in the general scenario.
\section{Applications}\label{5}
In the subsequent discussion, second-order linear differential equations frequently encountered in physics are addressed and resolved utilizing the recursion method based on generalized recursive integrating factors derived in the preceding section.
\subsection{Constant Coefficients Equation}
The constant coefficients equation (see \cite{zill2016differential}) in canonical form is given as follows:
\begin{equation} \label{cc 1}
    \frac{d^2y}{dx^2}+\frac{b}{a}\frac{dy}{dx}+\frac{c}{a}y(x)=f(x)
.\end{equation}
where $a\neq0$.The next step involves computing the $\beta(x)$ and $h(x)$ functions to determine $\alpha(x)$:

\begin{equation}\label{cc 2}
    \beta(x) = \exp\left( \int_0^x \frac{p(t)}{2}dt\right) = \exp \left(\frac{b}{2a}x \right)
.\end{equation}

\begin{equation} \label{cc 3}
    h(x)=-\frac{c}{a}+\frac{b^2}{4a^2}=\frac{b^2-4ac}{4a^2}
.\end{equation}
Since Eq.\eqref{cc 3} is constant, the computation for $\alpha(x)$ simplifies directly into the subsequent expansion:
\begin{equation} \label{cc 4}
    \alpha(x)=1+h\frac{x^2}{2!}+h^2 \frac{x^4}{4!}+...
\end{equation}
This corresponds precisely to the Taylor series expansion of $\exp({\sqrt{h}x})+\exp({-\sqrt{h}x})$ and is expressed as follows:
\begin{equation} \label{cc 5}
    \alpha(x)=\exp\left({\frac{\sqrt{b^2-4ac}}{2a}x}\right)+\exp \left({-\frac{\sqrt{b^2-4ac}}{2a}x}\right)=2\cosh\left(\frac{\sqrt{b^2-4ac}}{2a}x\right)
.\end{equation}
Upon determining $\alpha(x),\beta(x)$, it is possible to proceed to reconstruct the general solution for $y(x)$. It is necessary to evaluate the terms present in Eq.\eqref{eq 2.1.13} to derive $y(x)$:
\begin{equation} \label{cc 6}
    \frac{\alpha(x)}{\beta(x)} = \exp \left(\frac{-b+\sqrt{b^2-4ac}}{2a} x\right)+\exp \left(\frac{-b-\sqrt{b^2-4ac}}{2a} x\right)
.\end{equation}
\begin{equation}\label{cc 7}
    \int_0^x \frac{1}{\cosh^2\left(\frac{\sqrt{b^2-4ac}}{2a}x\right)} = \frac{2a}{\sqrt{b^2-4ac}}\tanh \left(\frac{\sqrt{b^2-4ac}}{2a}x\right) 
.\end{equation}
As evident from the general expression derived for $y(x)$ in Eq.\eqref{eq 2.1.13}, and in alignment with the general theory of linear differential equations \cite{zill2016differential,arfken2013mathematical}, the general solution for a second-order linear differential equation is constituted by the sum of a homogeneous solution $y_h(x)$ and a particular solution $y_p(x)$ for the inhomogeneous scenario:
\begin{equation}\label{extra2}
    y(x)=y_h(x)+y_p(x)
.\end{equation}
Based on Eq.\eqref{eq 2.1.13}, it is possible to derive the explicit expressions for $y_h(x),y_p(x)$ as follows:
\begin{equation}\label{ccextra1}
    y_h(x)=C_2\frac{\alpha(x)}{\beta(x)}+C_1\frac{\alpha(x)}{\beta(x)}\int_0^x \frac{dt}{\alpha^2(t)}
.\end{equation}
\begin{equation}\label{ccextra2}
    y_p(x)=\frac{\alpha(x)}{\beta(x)}\int_0^x \frac{1}{\alpha^2(t)}\int_0^t \alpha(t')\beta(t')f(t')dt' dt
.\end{equation}
\subsubsection{Homogeneous solution}
The homogeneous solution for the constant coefficient equation is derived by substituting the outcomes of Eq.\eqref{cc 6} and Eq.\eqref{cc 7} into Eq.\eqref{ccextra1}:
\begin{equation}\label{cc extra1}
    y_h(x)=C_1\exp\left(\frac{-b+\sqrt{b^2-4ac}}{2a}x\right)+C_2\exp\left(\frac{-b-\sqrt{b^2-4ac}}{2a}x\right)
.\end{equation}
This represents the exact general homogeneous solution to the constant coefficient equation as documented in \cite{zill2016differential,arfken2013mathematical} .
\subsubsection{Inhomogeneous solution}
The inhomogeneous solution for the constant coefficient equation is obtained by substituting the outcomes of Eq.\eqref{cc 6} and Eq.\eqref{cc 7} into Eq.\eqref{ccextra2}:
\begin{equation}
    y_p(x)=\frac{8a}{b^2-4ac} e^{ -\frac{b}{2a}x}\cosh \left(\frac{\sqrt{b^2-4ac}}{2a}x \right) \int_0^x \tanh \left( \frac{\sqrt{b^2-4ac}}{2a}t \right) \int_0^t e^{\frac{b}{2a}t'}\cosh \left(\frac{\sqrt{b^2-4ac}}{2a}t' \right)f(t')dt'dt
.\end{equation}
Contrary to conventional methodologies employed to determine the inhomogeneous component of the solution for $y(x)$, which often necessitate conjecturing the solution form (refer to \cite{zill2016differential} for the method of undetermined coefficients) or computing Wronskians (as illustrated in the variation of parameters technique in \cite{zill2016differential}), the present calculation is executed with relative ease. Consequently, the comprehensive solution for $y(x)$ is articulated as follows:
\begin{equation}\label{cc 8}
\begin{split}
y(x) =& C_1e^{ \left(\frac{-b+\sqrt{b^2-4ac}}{2a} x\right)}+C_2 e^{ \left(\frac{-b-\sqrt{b^2-4ac}}{2a} x\right)}\\ &+
\frac{8a}{b^2-4ac} e^{ -\frac{b}{2a}x}\cosh \left(\frac{\sqrt{b^2-4ac}}{2a}x \right) \int_0^x \tanh \left( \frac{\sqrt{b^2-4ac}}{2a}t \right) \int_0^t e^{\frac{b}{2a}t'}\cosh \left(\frac{\sqrt{b^2-4ac}}{2a}t' \right)f(t')dt'dt.
\end{split}
\end{equation}

\subsection{Cauchy-Euler equation}
The Cauchy-Euler equation as given in \cite{arfken2013mathematical,zill2016differential} is a second order linear differential equation in canonical form given by:
\begin{equation} \label{ce 1}
    \frac{d^2y}{dx^2}+\frac{b}{ax}\frac{dy}{dx}+\frac{c}{ax^2}=\frac{f(x)}{x^2}
.\end{equation}
where $a\neq0$.From Eq.\eqref{ce 1}the functions $p(x),q(x)$ are identified by:
\begin{equation} \label{ce 2}
    p(x)=\frac{b}{a}\frac{1}{x}, \quad q(x)=\frac{c}{a}\frac{1}{x^2}, \quad \frac{1}{2}\frac{dp}{dx}=-\frac{b}{2a}\frac{1}{x^2}
.\end{equation}
Equation \eqref{ce 2} facilitates the computation of expressions $\beta(x)$ and $h(x)$:
\begin{equation}\label{ce 3}
    \beta(x)=\exp\left({\int_0^x \frac{1}{2}\frac{b}{at}dt}\right)=x^{\frac{b}{2a}}
.\end{equation}
\begin{equation} \label{ce 4}
    h(x)=-q(x)+\frac{1}{2}\frac{dp}{dx}+\frac{p^2}{4}=-\frac{b}{2ax^2}+\frac{b^2-4ac}{4a^2x^2} = \frac{\gamma}{x^2}
.\end{equation}
where:
\begin{equation}\label{ce 5}
    \gamma = -\frac{b}{2a}+\frac{b^2-4ac}{4a^2}
.\end{equation}
Equation \eqref{ce 4} allows for the recursive computation of $\alpha(x)$:
\begin{equation}
    \alpha(x)=1-\gamma \ln(x)+\int_0^x\int_0^{t_1}\int_0^{t_2}\int_0^{t_3} \frac{1}{t_4^2t_2^2}dt_4dt_3dt_2dt_1+\hdots
\end{equation}
Computing the recursion:
\begin{dmath}\label{ce 6}
    \alpha(x)=1+\ln(x)\left[-\gamma+\gamma^2-2\gamma^3+5\gamma^4+\hdots \right]+\frac{\ln^2(x)}{2!}\left[\gamma^2-2\gamma^3+5\gamma^4+\hdots \right]+
    \frac{\ln^3(x)}{3!}\left[-\gamma^3+3\gamma^4-9\gamma^5+28\gamma^6 +\hdots\right]+\hdots
\end{dmath}
But:
\begin{equation}\label{ce 7}
    -\gamma+\gamma^2-2\gamma^3+5\gamma^4+\hdots=\frac{1}{2}-\frac{1}{2}\sqrt{1+4\gamma}
.\end{equation}
\begin{equation}\label{ce 8}
    \gamma^2-2\gamma^3+5\gamma^4+\hdots=\left[\frac{1}{2}-\frac{1}{2}\sqrt{1+4\gamma} \right]^2
.\end{equation}
\begin{equation}\label{ce 9}
    -\gamma^3+3\gamma^4-9\gamma^5+28\gamma^6 +\hdots=\left[\frac{1}{2}-\frac{1}{2}\sqrt{1+4\gamma} \right]^3
.\end{equation}
Consequently, Eq.\eqref{ce 6} can be simplified to the following expression:\begin{equation} \label{ce 10}
    \alpha(x) = \sum_{j=0}^{\infty} \frac{1}{j!}\left[\frac{1}{2}-\frac{1}{2}\sqrt{1+4\gamma} \right]^j ln^j(x)
.\end{equation}
This indeed represents the Taylor expansion series of the subsequent exponential function:
\begin{equation}\label{ce 11}
    \alpha(x)=\exp \left({\left[\frac{1}{2}-\frac{1}{2}\sqrt{1+4\gamma} \right]\ln(x)}\right)=x^{\left[\frac{1}{2}-\frac{1}{2}\sqrt{1+4\gamma} \right]}
.\end{equation}
Upon substituting the explicit value for $\gamma$, one obtains:
\begin{equation}
    1+4\gamma = \frac{(b-a)^2+4ac}{a^2}
.\end{equation}
Subsequently, the expression for $\alpha(x)$ is simplified to:
\begin{equation}\label{ce 12}
    \alpha(x)=x^{\frac{a-\sqrt{(b-a)^2+4ac}}{2a}}
.\end{equation}
Upon calculation of $\alpha(x),\beta(x)$, the development of the comprehensive solution $y(x)$ can begin by evaluating the terms that emerge in Eq.\eqref{eq 2.1.13}:
\begin{equation}\label{ce 13}
    \frac{\alpha(x)}{\beta(x)}=x^{{\frac{a-b-\sqrt{(b-a)^2+4ac}}{2a}}}
.\end{equation}
\begin{equation}\label{ce extra2}
    \frac{1}{\alpha^2(t)}=t^{{\frac{-a+\sqrt{(b-a)^2+4ac}}{a}}}
.\end{equation}
\begin{equation}\label{ce extra1}
\int_0^x \frac{dt}{\alpha(t)^2} = \left\{ 
\begin{array}{ll}
      Cx^{\frac{\sqrt{(b-a)^2+4ac}}{a}} & (b-a)^2+4ac\neq 0 \\
      \ln(x) & (b-a)^2+4ac=0
\end{array} 
\right. 
.\end{equation}
\subsubsection{Homogeneous solution}
The homogeneous solution for the Cauchy-Euler Equation is derived by utilizing Eq.\eqref{ce 13} through Eq.\eqref{ce extra1} into Eq.\eqref{ccextra1}:
\begin{equation}\label{ce extra3}
y_h(x)=  \left\{ 
\begin{array}{ll}
      C_1x^{{\frac{a-b-\sqrt{(b-a)^2+4ac}}{2a}}} +C_2x^{{\frac{a-b+\sqrt{(b-a)^2+4ac}}{2a}}} & (b-a)^2+4ac\neq 0 \\
      C_1x^{{\frac{a-b}{2a}}}+C_2x^{{\frac{a-b}{2a}}}\ln(x) & (b-a)^2+4ac=0
\end{array} 
\right. 
.\end{equation}
This solution is consistent with the general solution of the homogeneous Cauchy-Euler equation as documented in \cite{zill2016differential}.
\subsubsection{Inhomogeneous solution}
The particular solution $y_p(x)$, as determined by Eq.\eqref{ccextra2} after calculating $\alpha(x),\beta(x)$, is expressed as follows:
\begin{equation}\label{ce extra4}
    y_p(x) =x^{\frac{a-b-\sqrt{(b-a)^2+4ac}}{2a}}\int_0^x\int_0^t t_1^{\frac{a+b-\sqrt{(b-a)^2+4ac}}{2a}}t^{\frac{-a+\sqrt{(b-a)^2+4ac}}{a}}f(t_1)dt_1dt
.\end{equation}
Contrary to traditional approaches, the particular solution is derived in a straightforward manner.
\subsection{Airy Equation}
The Airy differential equation discovered by Airty in \cite{airy1838intensity} is given by:
\begin{equation} \label{airy 1}
\frac{d^2y}{dx^2}-xy=f(x)
.\end{equation}
From here $p(x),q(x)$ are identified by:
\begin{equation}\label{airy 2}
    p(x)=0,\quad q(x)=-x
.\end{equation}
Then the expressions for $h(x),\beta(x)$ are given by:
\begin{equation}\label{airy 3}
    h(x)=x \quad \beta(x)=1
.\end{equation}
Solving by recursion for $\alpha(x)$ it is obtained the following:
\begin{equation}\label{airy 4}
    \alpha(x)=1+\frac{x^3}{6}+\frac{x^6}{180}+\frac{x^9}{12960}+\hdots
\end{equation}
A pattern is observed from Eq.\eqref{airy 4}:
\begin{equation}\label{airy 5}
    \alpha(x)=1+\sum_{n=1}^{\infty}\frac{x^{3n}}{(3n+1)(3n)(3n-2)(3n-3)...3.2}
.\end{equation}
However, this precisely corresponds to the series expansion for the Airy function of the first kind in \cite{airy1838intensity}:

\begin{equation}\label{airy 6}
    \alpha(x)=Ai(x)
.\end{equation}
\subsubsection{Homogeneous solution}
According to $\beta(x)=1$, Equation \eqref{airy 6} represents one of the independent homogeneous solutions. In order to obtain the additional independent solution from Equation \eqref{ccextra1}, it is necessary to evaluate the integral:
\begin{equation}
    \int_0^{x}\frac{dt}{\alpha(t)^2} = x+\frac{x^4}{12}+\frac{13x^7}{1260}+\hdots
\end{equation}
Then:
\begin{equation}\label{airy 7}
    \frac{\alpha(x)}{\beta(x)}\int_0^{x}\frac{dt}{\alpha(t)^2}=1+\frac{x^4}{12}+\frac{x^7}{504}+\frac{x^{10}}{45360}+\hdots
\end{equation}
A recurrence pattern is identified:
\begin{equation}\label{airy 8}
     \frac{\alpha(x)}{\beta(x)}\int_0^{x}\frac{dt}{\alpha(t)^2}=x+\sum_{n=1}^{\infty}\frac{x^{3n+1}}{(3n+1)(3n)(3n-2)(3n-3)...(4)(3)}
.\end{equation}
This, however, precisely corresponds to the series expansion for the Airy function of the second kind:
\begin{equation}\label{airy 9}
     \frac{\alpha(x)}{\beta(x)}\int_0^{x}\frac{dt}{\alpha(t)^2}=Bi(x)
.\end{equation}
Consequently, the general homogeneous solution is expressed as:\begin{equation}\label{airy 10}
    y_h(x)=C_1 Ai(x)+C_2Bi(x)
.\end{equation}

\subsubsection{Inhomogeneous section}
The particular solution for the inhomogeneous case may be derived from Equation \eqref{ccextra2}:
\begin{equation}
    y_p(x)=Ai(x)\int_0^x \int_0^t \frac{Ai(t_1)}{Ai(t)^2}f(t_1)dt_1dt
.\end{equation}
Contrary to conventional approaches, the particular solution is directly derived.
\subsection{Legendre Equation}
The Legendre equation discovered by Legendre in \cite{legendre1790recherches} is given by:
\begin{equation}\label{legendre 1}
    \frac{d^2y}{dx^2}-\frac{2x}{1-x^2}\frac{dy}{dx}+\frac{l(l+1)}{1-x^2}y(x)=f(x)
.\end{equation}
From this equation $p(x),h(x)$ are identified by:
\begin{equation}\label{legendre 2}
    p(x)=-\frac{2x}{1-x^2}, \quad h(x)=-\frac{l(l+1)+1}{1-x^2}-\frac{x^2}{(1-x^2)^2}.
\end{equation}
Then $\beta(x)$ is computed:
\begin{equation}\label{legendre 3}
    \beta(x)=\exp\left({\frac{1}{2}\int_0^x p(t)dt}\right)=\sqrt{1-x^2}
.\end{equation}
Following this, the derivation for $\alpha(x)$ is computed:
\begin{equation}\label{legendre 4}
    \alpha(x)=1+-\frac{l(l+1)+1}{2}x^2+\frac{l(l+1)^2-3}{24}x^4+\hdots
\end{equation}
\subsubsection{Homogeneous solution}
Upon completion of the computation in $\alpha(x),\beta(x)$, the initial homogeneous solution can be constructed as follows:
\begin{equation}\label{legendre 5}
    y_1(x)=\frac{\alpha(x)}{\beta(x)}=1-\frac{l(l+1)}{2!}x^2+\frac{(l-2)(l)(l+1)(l+3)}{4!}x^4+\hdots
\end{equation}
The identified pattern emerging from the expansion indicates that $y_1(x)$ simplifies to:
\begin{equation}\label{legendre 6}
    y_1(x)=1+\sum_{n=1}^{\infty}(-1)^n \frac{(l-2n+2)...(l-2)(l)(l+1)(l+3)...(l+2n-1)}{(2n)!}x^{2n}
.\end{equation}
Upon completing the computation detailed in $y_1(x)$, it becomes feasible to derive the second homogeneous solution:
\begin{equation}\label{legendre 7}
    \int_0^x \frac{dt}{\alpha(t)^2}=x+\frac{1+l+l^2}{3}x^3+\left(\frac{1}{5}+\frac{3}{10}l+\frac{13}{30}l^2+\frac{4}{15}l^3 +\frac{2}{15}l^4\right)x^5+\hdots
\end{equation}
Then:
\begin{equation}\label{legendre 8}
    y_2(x)=y_1(x)\int_0^x\frac{dt}{\alpha^2(t)}=x-\frac{(1-l)(2+l)}{3!}x^3+\frac{(l-3)(l-1)(l+2)(l+4)}{5!}x^5+\hdots
\end{equation}
At this point, a pattern within the recursive process is discerned, whereby $y_2(x)$ simplifies to:
\begin{equation}\label{legendre extra1}
    y_2(x)=x+\sum_{n=0}^{\infty}(-1)^n\frac{(l-2n+1)...(l-1)(l+2)(l+4)...(l+2n)}{(2n+1)!}x^{2n+1}
.\end{equation}
Consequently, the general homogeneous solution is articulated as follows:
\begin{equation}\label{Legendre extra2}
    y_h(x)=C_1y_1(x)+C_2y_2(x)
.\end{equation}
When $l$ is an even integer, $y_1(x)$ is reduced to a Legendre polynomial\cite{legendre1790recherches} of degree $l$ $P_l(x)$. Meanwhile, $y_2(x)$ exhibits divergence at $x=\pm1$ and is referred to as the Legendre function of the second kind $Q_l(x)$. Conversely, when $l$ is an odd integer, $y_2(x)$ is reduced to a Legendre polynomial of degree $l$ $P_l(x)$, and $y_1(x)$ demonstrates divergence at $x=\pm1$, being similarly identified as the Legendre function of the second kind $Q_l(x)$. Accordingly, the homogeneous solution presented in Eq.\eqref{Legendre extra2} is expressed for $l$ as an integer as follows:
\begin{equation}\label{Legendre extra3}
    y_h(x)=C_1P_l(x)+C_2Q_l(x)
.\end{equation}
\subsubsection{Inhomogeneous solution}
The computation for the particular solution $y_p(x)$ is directly derived from Equation \eqref{ccextra2} in a straightforward manner:
\begin{equation}
    y_p(x)=y_1(x)\int_0^x\int_0^t \frac{y_1(t_1)(1-t_1^2)}{y_1(t)^2(1-t^2)}f(t_1) dt_1 dt
.\end{equation}
\subsection{Hermite Equation}
The Hermite differential equation discoverd by Hermite in \cite{bhlpage3659507} is given by:
\begin{equation}\label{hermite 1}
    \frac{d^2y}{dx^2}-2x\frac{dy}{dx}+ay=f(x)
.\end{equation}
From Eq.\eqref{hermite 1} $p(x),h(x)$ are identified:
\begin{equation}\label{hermite 2}
    p(x)=-2x,\quad h(x)=x^2-a-1
.\end{equation}
Then $\beta(x)$ is given by:
\begin{equation}\label{hermite 3}
    \beta(x)=\exp \left({-\frac{x^2}{2}}\right)
.\end{equation}
The recursive solution for $\alpha(x)$ is obtained as follows:
\begin{equation}\label{hermite 4}
    \alpha(x)=1-\frac{1+a}{2!}x^2+\frac{a^2+2a+3}{4!}x^4+\frac{(a+1)(a^2+2a+15)}{6!}x^6+\hdots
\end{equation} 
\subsubsection{Homogeneous Solution}
Upon calculating $\alpha(x),\beta(x)$, one may begin the construction of the initial homogeneous solution followin Eq.\eqref{ccextra1}:
\begin{equation}\label{hermite 5}
    y_1(x)=\frac{\alpha(x)}{\beta(x)}=1-\frac{a}{2!}x^2-\frac{(4-a)a}{4!}x^4-\frac{(8-a)(4-a)a}{6!}x^6+\hdots
\end{equation}
The second homogeneous solution can be computed once again from Eq.\eqref{ccextra1}:
\begin{equation}\label{hermite 6}
    y_2(x)=\frac{\alpha(x)}{\beta(x)}\int_0^x \frac{dt}{\alpha(t)^2}=x+\frac{2-a}{3!}x^3+\frac{(6-a)(2-a)}{5!}x^5+\hdots
\end{equation}
The general form of the homogeneous solution is thus articulated as follows:
\begin{equation}
    y_h(x)=C_1y_1(x)+C_2y_2(x)
.\end{equation}

This aligns with the documented general homogeneous solution of Hermite's equation as documented in \cite{arfken2013mathematical}.
\subsubsection{Inhomogeneous solution}
The computation of the particular solution $y_p(x)$ proceeds directly from Eq.\eqref{ccextra2}:
\begin{equation}
    y_p(x)=y_1(x)\int_0^x\int_0^t \frac{y_1(t_1)\exp(-t_1^2)}{y_1(t)^2\exp(-t^2)}f(t_1) dt_1 dt
.\end{equation}
\subsection{Chebyshev Equation}
The Chebyshev differential equation discovered by Chebyshev in \cite{cheb} is given by:
\begin{equation}\label{cheb1}
    \frac{d^2y}{dx^2}-\frac{x}{1-x^2}\frac{dy}{dx}+a^2y=f(x)
.\end{equation}
From Eq.\eqref{cheb1} $p(x),h(x)$ are identified:
\begin{equation}\label{cheb2}
    \beta(x)=(1-x^2)^{\frac{1}{4}},\quad h(x)=-\frac{a^2}{1-x^2}-\frac{3x^2}{4(1-x^2)^2}-\frac{1}{2(1-x^2)}
.\end{equation}
Recursively solving for $\alpha(x)$ the following is obtained:
\begin{equation}\label{cheb3}
    \alpha(x)=1-\left(\frac{a^2}{2}+\frac{1}{4}\right)x^2+\left(\frac{a^4}{24}-\frac{a^2}{24}-\frac{3}{32}\right)x^4+\hdots
\end{equation}
\subsubsection{Homogeneous Solution}
The initial homogeneous solution of Eq.\eqref{cheb1} can be derived by employing Eq.\eqref{cheb2} and Eq.\eqref{cheb3}:
\begin{equation}\label{cheb5}
    y_1(x)=1-\frac{1}{2}a^2x^2+\left(\frac{a^4-4a^2}{24}\right)x^4+\hdots
\end{equation}
From Eq.\eqref{cheb5}, a discernible pattern emerges that facilitates the representation of $y_1(x)$ using compact notation:
\begin{equation}\label{cheb6}
    y_1(x)=\sum_{n:even}\frac{(2n^2-a^2)((2n-2)^2-a^2)\hdots(-a^2)}{(2n)!}a_0 x^{2n}
.\end{equation}
Following the computation of $y_1(x)$, the remaining linearly independent homogeneous solution is determined:
\begin{equation}\label{cheb7}
    \int_0^x \frac{dt}{\alpha(t)^2}=x+(\frac{a^2}{3}+\frac{1}{6})x^3+\hdots 
\end{equation}
Then:
\begin{equation}\label{cheb8}
    y_2(x)=x+\frac{1-a^2}{3!}x^3+\frac{(3^2-a^2)(1^2-a^2)}{5!}x^5+\hdots
\end{equation}
From Eq.\eqref{cheb8}, a discernible pattern is identified, which facilitates the expression of $y_2(x)$ in a more concise notation:
\begin{equation}\label{cheb9}
    y_2(x)=\sum_{n:odd} \frac{((2n-1)^2-a^2)((2n-3)^2-a^2)\hdots(1^2-a^2)}{(2n+1)!}a_0x^{2n+1}
.\end{equation}
As illustrated by Boyce in \cite{boyce2021elementary}, the coefficients of the series expansions present in Eq.\eqref{cheb7} and Eq.\eqref{cheb8} can be reformulated in a closed form as below:
\begin{equation}\label{cheb10}
\begin{split}
    &c_{k:odd}=\frac{2^{k-1}\pi a\sec(\frac{1}{2}\pi a)}{\Gamma(1-\frac{1}{2}k-\frac{1}{2}a)\Gamma(1-\frac{1}{2}k+\frac{1}{2}a)}a_0,\\
    &c_{k:even}=\frac{2^{k-1}\pi a\csc(\frac{1}{2}\pi a)}{\Gamma(1-\frac{1}{2}k-\frac{1}{2}a)\Gamma(1-\frac{1}{2}k+\frac{1}{2}a)}a_0.
\end{split}
\end{equation}
Then Eq.\eqref{cheb7} and Eq.\eqref{cheb8} according to Boyce\cite{boyce2021elementary} can be rewritten as:
\begin{equation}\label{cheb11}
    y_1(x)=\cos(a\sin^{-1}(x)),\quad y_2(x)=\sin(a\sin^{-1}x)
.\end{equation}
Thus, the general homogeneous solution can be expressed as:\begin{equation}\label{cheb110}
    y(x)=C_1\cos(a\sin^{-1}(x))+ C_2y_2(x)=\sin(a\sin^{-1}x)
.\end{equation}
This is, once more, consistent with the previously documented solutions\cite{arfken2013mathematical,boyce2021elementary} of Chebyshev's differential equation.
\subsubsection{Inhomogeneous solution}
From Eq.\eqref{eq 2.1.13} the particular solution $y_p(x)$ is given by:
\begin{equation}\label{cheb12}
    y_p(x)=y_1(x)\int_0^x\int_0^t \frac{y_1(t_1)(1-t_1)^{1/2}}{y_1(t)^2(1-t)^{1/2}}f(t_1) dt_1 dt
.\end{equation}

\section{Conclusions}\label{6}
An alternative approach has been introduced to address ordinary differential equations through recursion. The method, as explored in this study, has demonstrated its utility in finding the general solution to first-order linear differential equations, yielding results consistent with those obtained via the integrating factor method in \cite{zill2016differential}. Furthermore, it has proven effective in solving second-order linear differential equations under general conditions, as illustrated in Eq.\eqref{eq 2.1.13}, through a generalization of the integrating factors method, wherein the factors are determined recursively. This methodological framework has enabled the derivation of known solutions for various second-order equations pertinent to numerous physical phenomena. Consequently, it is concluded that the recursive method holds significant value in deriving general solutions for linear differential equations of both first and second orders.In the context of second-order differential equations, this method offers a systematic and logical framework for solving these equations, contrasting with the heuristic guessing approaches commonly available in existing literature. \\
As indicated in Section \ref{2}, the recursion method is not restricted to resolving linear differential equations; rather, it has the potential to be applicable to non-linear differential equations, provided that one can discern patterns within the recursion. Hence, a promising avenue for further investigation is the application of this method to non-linear differential equations, principally those encountered in the field of physics. A topic to be explored in future studies.


%
%

%

\begin{acknowledgments}
Everardo Rivera-Oliva wishes to express his gratitude for the support provided by the Ph.D. scholarship from SECIHTI (CONAHCYT)-Mexico.
\end{acknowledgments}
\bibliography{biblio}

\end{document}